\documentclass[10pt,letterpaper,onecolumn,oneside]{article}
\usepackage[top=0.85in,left=1.75in,footskip=0.75in,marginparwidth=2in]{geometry}

\usepackage{amsmath}
\usepackage{booktabs} 
\usepackage{nopageno}
\usepackage{rotating}

\newcommand{\paolo}[1]{\textcolor{black}{#1}}

\usepackage[utf8]{inputenc}

\usepackage[colorlinks]{hyperref}

\usepackage[right]{lineno}

\usepackage{microtype}
\DisableLigatures[f]{encoding = *, family = * }

\usepackage{changepage}

\usepackage[aboveskip=1pt,labelfont=bf,labelsep=period,singlelinecheck=off]{caption}

\makeatletter
\renewcommand{\@biblabel}[1]{\quad#1.}
\makeatother

\usepackage{graphicx}
\usepackage{amssymb}
\usepackage{amsmath}
\usepackage{comment}
\usepackage{enumerate}
\usepackage{color}
\usepackage{subeqnarray}
\usepackage{ulem}

\usepackage{lastpage,fancyhdr,graphicx}
\usepackage{color}

\definecolor{Gray}{gray}{.25}

\usepackage{graphicx}

\usepackage{sidecap}

\usepackage{wrapfig}
\usepackage[pscoord]{eso-pic}
\usepackage[fulladjust]{marginnote}
\reversemarginpar

\usepackage{comment}
\usepackage{siunitx}
\usepackage{textcomp}
\usepackage{units}
\usepackage{ulem}

\DeclareSIUnit\Molar{\textsc{m}} 

\begin{document}
\thispagestyle{plain}

\vspace*{0.35in}

\begin{flushleft}
{\Large
\textbf\newline{An entrainment model for fully-developed wind farms: effects of atmospheric stability and an ideal limit for wind farm performance}
}
\newline
\\
{Paolo Luzzatto-Fegiz}\textsuperscript{a}\footnote{ fegiz@engineering.ucsb.edu} and
{Colm-cille P. Caulfield}\textsuperscript{b,c}

\bigskip
\textsuperscript{a} Department of Mechanical Engineering, University of California, Santa Barbara, CA 93106, USA
\\
\textsuperscript{b} BP Institute for Multiphase Flow, Madingley Rise, Madingley Road, Cambridge, CB3 0EZ, UK
\\
\textsuperscript{c} Department of Applied Mathematics and Theoretical Physics, University of Cambridge, Wilberforce Rd, CB3 0WA, UK
\\

\end{flushleft}

\section*{Abstract}
While a theoretical limit has long been established for the performance of a single turbine, no corresponding upper bound exists for the power output from a large wind farm, making it difficult to evaluate the available potential for further performance gains. 
	{Recent work involving vertical-axis turbines has achieved large increases in power density relative to traditional wind farms (Dabiri, J.O., {\it J. Renew. Sust. Energy} {\bf 3}, 043104 (2011)), thereby adding motivation to the search for an upper bound.}
	Here we build a model describing the essential features of a large array of turbines with arbitrary design and layout, by considering a fully-developed wind farm whose upper edge is bounded by a self-similar boundary layer.
	The exchanges between the wind farm, the overlaying boundary layer, and the outer flow are parameterized by means of the classical entrainment hypothesis. We obtain a concise expression for the wind farm's power density (corresponding to power output per unit planform area), as a function of three coefficients, which represent the array thrust and the turbulent exchanges at each of the two interfaces.
	{Before seeking an upper bound on farm performance}, we assess the performance of our {simple model} by comparing the predicted power density to field data, \paolo{laboratory measurements and large-eddy simulations} for the fully-developed regions of wind farms, finding \paolo{good} agreement. 
	Furthermore, we extend our model to include the effect of atmospheric stability on power output, by using a parameterization {(which had been previously developed in the context of geophysical fluid dynamics)} relating entrainment coefficients to local Froude numbers. Our predictions for power variation with atmospheric stability are in {broad} agreement with field measurements. To the best of our knowledge, this constitutes the first quantitative comparison between an atmospheric-stability-dependent model and field data.
	Finally, we consider an ideal limit for array operation, whereby turbines are designed to maximize momentum exchange with the overlying boundary layer.
	This enables us to obtain an upper bound for the performance of large wind farms, which we determine to be an order of magnitude larger than the output of contemporary turbine arrays.

\vspace*{0.7in}


\section{Introduction}
\begin{figure}[t]	
\centering
\includegraphics[width=0.99\textwidth]{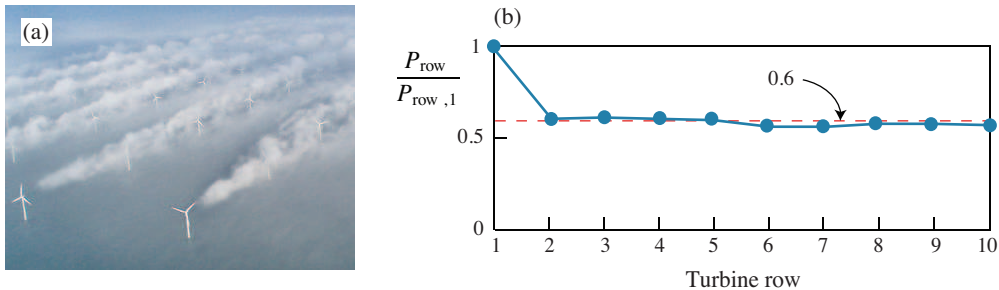}
	\caption{(a) Aerial view of the Horns Rev wind farm; turbine trails are visualized by condensation (photograph by Christian Steiness, adapted from Ref.~\cite{Hasager_etal_Energies_2013}). (b) Power produced by each turbine row, normalized by the power output from the first row, when the wind direction is aligned with the turbine grid \cite{Barthelmie_etal_JAOT_2010}.} 
    \label{fig:HornsRev}
\end{figure}

While it has been argued that wind energy presents a large untapped potential \cite{Lu_etal_PNAS_2009}, and that it could realistically mitigate climate change \cite{Barthelmie_Pryor_NCC_2014}, a persistent shortcoming of wind farms is that their power density (defined as power output per unit land area) is low, reaching typically only a few W/m$^2$ in a large farm \cite{Dabiri_JRSE_2011, Barthelmie_etal_JAOT_2010}. For this reason, the production of substantial amounts of electricity from wind requires the construction of extensive wind farms, which can comprise hundreds of units (an example is given by the {Horns Rev} facility, shown in Fig.~\ref{fig:HornsRev}A). The flow physics of these large turbine arrays are fundamentally different from those of a single turbine operating in isolation; in a large wind farm, most of the fluid kinetic energy enters not from the front, but rather through turbulent motions at the top of the array \cite{Frandsen_etal_WE_2006}. Remarkably, while a theoretical limit has long been established for the performance of a single turbine \cite{Betz_GN_1919,Sorensen_ARFM_2011}, no corresponding theory appears to exist for a general, large-scale energy extraction array. Recently, Ref.~\cite{Goit:2015cn} obtained numerical results showing that power increases of the order of 10\% may be possible, by applying optimal control techniques to traditional turbine designs. {Work with {novel layouts of} vertical-axis turbines has demonstrated large performance gains relative to conventional turbine arrays  \cite{Dabiri_JRSE_2011}. Somewhat surprisingly, the power density that has been measured is even larger than the value that would be achieved if a traditional wind farm could be operated without interference losses ({i.e.} without changes in turbine spacing). These results add urgency to the search for a theoretical upper bound.}

Previous work on theoretical modeling of turbine arrays has focused on two main methodologies. The most commonly used wind-farm theories are based on superposing empirical models for wakes of individual turbines  \cite{Frandsen_etal_WE_2006, Katic_etal_1986}. As such, these do not consider effects arising from nonlinear wake interactions, which can be significant in affecting momentum fluxes in large arrays (as is the case, for example, in vegetation canopies \cite{Nepf_ARFM_2012}). In sufficiently large wind farms, the wakes eventually merge laterally, and the flow is often approximated as a canonical turbulent boundary layer. The wind farm is treated by resorting to the classic concept of effective roughness \cite{Frandsen_etal_WE_2006, Calaf_etal_PF_2010}, whereby the wind farm drag is imposed through a local increase in the effective roughness height. This provides a concise approximation that can be readily implemented in atmospheric models, provided the new roughness height is known. However, this approach is constrained by the need for semi-empirical formulae relating the effective roughness height (which is different from the actual obstacle size) to the physical turbine design and layout \cite{Calaf_etal_PF_2010}. These limitations make current wind farm models unsuitable for examining arrays of energy-extraction devices of arbitrary design and layout.

A further issue that has received attention involves the thermal coupling between the atmosphere and the array. While much work has focused on the effect of large wind farms on surface temperatures \cite{Keith_etal_PNAS_2004}, recent observations have shown that power output is increased when the atmosphere transitions from a stable condition to a neutral or unstable state (for a given wind velocity \cite{Hansen_etal_WE_2011, Barthelmie_etal_IEEE_2013}). 
Modeling of these atmospheric stability effects has focused on extending classical effective-roughness models (by making use of Monin-Obukhov similarity theory \cite{Emeis_WE_2010, Pena_Rathmann_WE_2014}), and on estimating corresponding corrections for models describing individual wakes \cite{Pena_Rathmann_WE_2014}. However, to the best of our knowledge, there are no investigations of atmospheric stability effects that have provided a quantitative comparison between theory and field data.

In this paper, we develop an entrainment-based model for power output from a large wind farm, and employ this theory to address the outstanding questions raised above. In Sec.~\ref{sec:model}, we introduce our entrainment-based wind farm model, and compare its predictions to field measurements from full-scale wind farms. In Sec.~\ref{sec:stab}, we extend the model to encompass the effect of atmospheric stability, and test its predictions against available full-scale data. In Sec.~\ref{sec:limit} we consider the limit of an `ideal' wind farm. The results are discussed briefly in Sec.~\ref{sec:disc}. Finally, conclusions are presented in Sec.~\ref{sec:conclusions}.

\section{An entrainment model for large wind farms}\label{sec:model}
{
	In this section, we employ the classical entrainment hypothesis {(generally attributed to G.~I.~Taylor, see e.g. Refs.~\cite{Morton_etal_PRS-A_1956},~\cite{Ellison_Turner_JFM_1959})} to develop a model for flow in a fully-developed canopy. Since the entrainment hypothesis has not previously been used in the context of canopy flows or wind farm aerodynamics (to the best of our knowledge), the next subsection briefly introduces the key elements of this turbulence closure.
	
	\subsection{The entrainment hypothesis, and its implications for boundary layers}
	The fundamental assumption {underlying the entrainment hypothesis} concerns the rate at which the interface between turbulent and nonturbulent fluid advances into the nonturbulent region. While the assumption is usually discussed in the context of plumes or gravity currents, here we consider a turbulent boundary layer, with outer velocity $U_o$. The (Reynolds-averaged) velocity $\overline{w}_E$ at which low-turbulence fluid crosses into the turbulent interface, measured in a frame of reference moving with the interface, is assumed to be \cite{Morton_etal_PRS-A_1956, Ellison_Turner_JFM_1959}:
	\begin{equation}\label{eq:E}
	\overline{w}_E = -E|U_b - U_o|,
	\end{equation}
	where the minus sign corresponds to the fact that turbulent fluid is below the low-turbulence fluid, in this example. Here, $U_b$ is a characteristic velocity for the interior of the boundary layer (defined below), an overline denotes Reynolds averaging, and $E$ is the entrainment coefficient, which is an empirical, nondimensional parameter. Much effort has been devoted to measuring $E$ in laboratory experiments, as well as in the ocean and atmosphere, as a function of Reynolds number and ambient stratification \cite{Cenedese_Adduce_JPO_2010}. In this parameterization, $U_b$ is defined by the relations \cite{Ellison_Turner_JFM_1959}
	\begin{eqnarray}
	(U_b - U_o)h_b &=& \int_0^{z_o} (\bar{u}-U_o) \, dz, \label{eq:def1} \\
	(U_b - U_o)^2h_b &=& \int_0^{z_o} (\bar{u}-U_o)^2 \, dz,
	\end{eqnarray}
	where $\bar{u}$ is the Reynolds-averaged $x$-component of velocity, $z$ is the upward-pointing direction, and the upper bound of integration $H$ is chosen so that $\overline{u}(z=z_o) = U_o$. { One should choose $z_o$ to be large enough to be outside the wind farm's boundary layer, but not so large that the resulting $U_o$ is no longer representative of the flow velocity driving the wind farm. Note that, in general, the value of $U_o$ will depend on atmospheric conditions.} To obtain a corresponding set of depth-integrated equations, which can incorporate (\ref{eq:E}), we start from the Reynolds-averaged mass and momentum equations, assuming the flow is steady and statistically independent of the transverse direction $y$, and retaining only the leading Reynolds and viscous stresses:
	\begin{eqnarray}
	\frac{\partial\overline{u}}{\partial x} +  \frac{\partial\overline{w}}{\partial z} &=& 0 \label{eq:RAcont}\\
	\frac{\partial\overline{u}^2}{\partial x} + \frac{\partial\overline{u}\,\overline{w}}{\partial z}
	&=& \frac{\partial}{\partial z}
	\left(- \overline{u'w'} + \nu \frac{\partial \overline{u}}{\partial z}\right) \label{eq:RAmom},
	\end{eqnarray}
	where the Reynolds decomposition for each quantity is denoted by $u = \overline{u} + u'$, {such that $\overline{u}$ is the ensemble-average (often implemented as a time-average) and $u'$ denotes a fluctuation from the ensemble, satisfying $\overline{u'}=0$.} In~(\ref{eq:RAmom}), $\nu$ is the kinematic viscosity and we consider a zero pressure gradient flow. We integrate (\ref{eq:RAcont}) from $z=0$ to $z=z_o$ to obtain
	\begin{eqnarray}
	\int_0^H \frac{\partial \overline{u}}{\partial x} dz + \left. \overline{w} \right|_{z={z_o}}-\left.\overline{w}\right|_{z=0} &=&   \frac{d}{d x}\int_0^{z_o} (\overline{u}-U_o) dz - (\left. \overline{u} \right|_{z={z_o}}-U_o)\frac{dH}{dx} + \left.\overline{w}\right|_{z={z_o}} \nonumber \\
	&=&\frac{d}{d x} \left[ h_b (U_b-U_o) \right] + \left.\overline{w}\right|_{z={z_o}} = 0 \label{eq:blah}
	\end{eqnarray}
	where we have used the no-through-flow condition at $z=0$, and have assumed ${\partial U_o}/{\partial x} = 0$ and $\left.\overline{u'w'}\right|_{z={z_o}} = 0$. We rearrange~(\ref{eq:blah}) to obtain
	\begin{equation}\label{eq:blah2}
	\frac{d}{d x} \left( h_b U_b \right) = -\left(\left.\overline{w}\right|_{z={z_o}}-  U_o\frac{d h_b}{d x}\right).
	\end{equation}
	Note that, by the chain rule of differentiation, $U_o\frac{d h_b}{d x} = \frac{d h_b}{d t}$ is the vertical velocity of the turbulent interface. Therefore the term in parentheses on the right-hand-side of (\ref{eq:blah2}) is the vertical fluid velocity in a frame of reference moving with the turbulent interface, that is, $\overline{w}_E$. Using (\ref{eq:E}), we obtain the integral form of the continuity equation {(\ref{eq:RAcont})}, under the entrainment hypothesis:
	\begin{equation}\label{eq:blMass}
	\frac{d}{d x} \left( h_b U_b \right) = E |U_o-U_b| .
	\end{equation}
	Similarly, we integrate the momentum equation {(\ref{eq:RAmom})} as follows. Note that the first term on the left-hand-side of~(\ref{eq:RAmom}) becomes
	\begin{eqnarray}
	\int_0^{z_o} \frac{\partial \overline{u}^2}{\partial x} \, dz &=& \int_0^{z_o} \left[  \frac{\partial (\overline{u}-U_o)^2}{\partial x}  + 2U_o \frac{\partial (\overline{u}-U_o) }{\partial x} \right] \, dz \nonumber \\
	&=& \frac{d}{dx} \int_0^{z_o} (\overline{u}-U_o)^2 \,dz + 2U_o  \frac{d}{dx} \int_0^{z_o} (\overline{u}-U_o) \, dz \nonumber \\
	&=& \frac{d}{dx} [h_b (U_b-U_o )^2] + 2U_o  \frac{d}{dx} [h_b(U_b-U_o)],
	\end{eqnarray}
	such that the momentum equation is
	\begin{equation}
	\frac{d}{dx} [h_b (U_b-U_o )^2] + 2U_o  \frac{d}{dx} [h_b(U_b-U_o)] = -U_o \left.\overline{w}\right|_{z={z_o}} - \nu \left.\frac{\partial u}{\partial z}\right|_{z=0}.
	\end{equation}
	Expanding the terms in the left-hand side and rearranging we have
	\begin{equation}
	\frac{d}{dx} [h_b U_b^2] = -U_o \left(\left.\overline{w}\right|_{z={z_o}}- U_o\frac{d h_b}{d x} \right) -\nu \left.\frac{\partial u}{\partial z}\right|_{z=0}.
	\end{equation}
	Using the definition of the entrainment velocity, we finally obtain the integral version of the momentum equation under the entrainment hypothesis:
	\begin{equation}\label{eq:intMomBL}
	\frac{d}{dx} [h_b U_b^2] = E \, U_o |U_o-U_b| -\nu \left.\frac{\partial u}{\partial z}\right|_{z=0}.
	\end{equation}
	When discussing stresses in turbulent boundary layers, it is common to consider the idealized case of a flow that is very slowly evolving in the $x$-direction \cite{Pope_CUP_2000}, such that the total stress (from turbulent fluctuations and viscosity) is constant. In such a case, (\ref{eq:intMomBL}) implies $E \, U_o |U_o-U_b| \simeq\nu \left.\frac{\partial u}{\partial z}\right|_{z=0} = \tau$, where $\tau$ is the wall stress. Since this is equal to the magnitude of the Reynolds stress in the boundary layer, we have {shown} that the first term on the right-hand-side of (\ref{eq:intMomBL}) corresponds to the Reynolds stress.
	
	\subsection{A two-interface entrainment model for fully-developed flow in wind farms}
	We distinguish three horizontal regions involving qualitatively different dynamics, as shown in Fig.~\ref{fig:FarmSchematic}. Starting from the bottom, these layers comprise the wind farm, a boundary layer and an outer region far above the farm. This implies that fluxes are governed by processes at two interfaces, which we label as the `farm' and `outer' interfaces. We will assume that the properties of the upper interface are primarily determined by atmospheric conditions, whereas those of the lower interface are directly affected by the wind farm design, as described below.
	
	To model the flow through the farm, we consider the double-averaged canopy equations, in a zero pressure gradient, where we retain only the leading-order stresses \cite{Finnigan_ARFM_2000,Nikora_etal_JHE_2007}
	\begin{eqnarray}
	\frac{\partial\langle\overline{u}\rangle}{\partial x} +  \frac{\partial\langle\overline{w}\rangle}{\partial z} &=& 0 \label{eq:DAcont}\\
	\frac{\partial\langle\overline{u} \rangle^2}{\partial x} + \frac{\partial\langle\overline{u}\rangle\langle\overline{w}\rangle}{\partial z}
	&=& \frac{\partial}{\partial z}
	\left(- \langle\overline{u'w'}\rangle - \langle{\overline{u}''\overline{w}''}\rangle + \nu \frac{\partial \langle\overline{u}\rangle}{\partial z}\right) + f \label{eq:DAmom}.
	\end{eqnarray}
	{These equations can be obtained, for example, by starting with the ensemble-averaged equations, and further decomposing $\overline u = \langle \overline u \rangle + \overline{u}''$, where the} angle brackets denote an horizontal average over one canopy wavelength in $x$ and $y$, that is
	\begin{equation}
	\langle u\rangle(x,y,z) = \frac{1}{\lambda_x \lambda_y}  \int_{x-\lambda_x/2}^{x+\lambda_x/2} \int_{y-\lambda_y/2}^{y+\lambda_y/2} u(x',y',z) \; dx' \, dy',
	\end{equation}
	{and $\overline{u}''$ is defined such that $\langle \overline{u}'' \rangle = 0$. Substituting $\overline u = \langle \overline {u} \rangle + \overline{u}''$ into (\ref{eq:RAcont}, \ref{eq:RAmom}), and averaging over the canopy wavelength, one obtains (\ref{eq:DAcont}, \ref{eq:DAmom}). The resulting dispersive stress $\langle{\overline{u}''\overline{w}''}\rangle$ accounts for exchanges of momentum due to flow features smaller than the spacing between canopy elements.} In the momentum equation (\ref{eq:DAmom}), $f$ is a body force corresponding to the drag exerted by the canopy, per unit mass.
	
	\begin{figure}[t]
    \centering
		\includegraphics{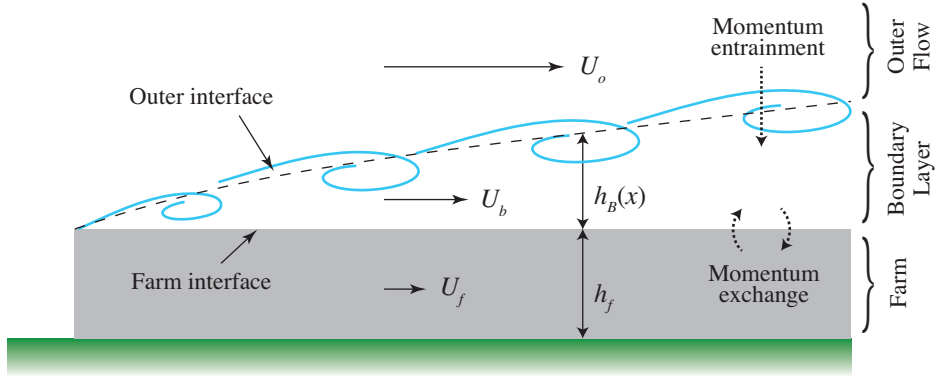}
        \vspace{10pt}
		\caption{Sketch of our entrainment model for a wind farm (the boundary layer is not to scale).} \label{fig:FarmSchematic}
	\end{figure}

	We consider the fully-developed regime, such that the velocity in the farm is independent of downstream distance $x$. This assumption is consistent with power measurements in large wind farms \cite{Barthelmie_etal_JAOT_2010, Hansen_etal_WE_2011}, as also shown in Fig.~\ref{fig:HornsRev}(b), and corresponds, inside the farm, to $\partial\langle\overline{u}\rangle/\partial x = 0$. By continuity, this implies $\partial\langle\overline{w}\rangle/\partial z = 0$. Since $\langle \overline{w} \rangle=0$ at the wall, we must have $\langle \overline{w} \rangle=0$ throughout the wind farm. Then the momentum equation (\ref{eq:DAmom}) is
	\begin{equation}
	0=\frac{\partial}{\partial z}
	\left(- \langle\overline{u'w'}\rangle - \langle\overline{u}''\overline{w}''\rangle + \nu \frac{\partial \langle\overline{u}\rangle}{\partial z}\right) + f.
	\end{equation}
	Integrating this from $z = 0$ to $z = h_f$, where $h_f$ is height of the wind farm:
	\begin{equation}
	\left(\langle\overline{u'w'}\rangle +\langle\overline{u}''\overline{w}''\rangle \right)_{z=h_f} = \int_0^{h_f} f \,dz  -\nu \left. \frac{\partial \langle\overline{u}\rangle}{\partial z} \right|_{z=0}.
	\end{equation}
	We now wish to express the stress and force terms in terms of depth-averaged velocities. To this end, let us define the canopy velocity as
	\begin{equation}\label{eq:vertAverage}
	h_f U_f = \int_0^{h_f} \langle \overline{u} \rangle \,dz.
	\end{equation}
	Assuming $U_f>0$, the vertically-integrated wind farm drag is written as (this step is similar to existing theories for vegetation canopies, and to other wind farm models \cite{Irwin_AE_1979, Nepf_ARFM_2012, Frandsen_etal_WE_2006})
	\begin{equation}
	\int_0^{h_f} f \; dz = -\frac{c_{ft}'}{2} U_f^2 
	\end{equation}
	{The coefficient $c_{ft}'$ accounts for the turbines' individual thrust and layout, and is defined as \cite{Finnigan_ARFM_2000, Calaf_etal_PF_2010}
		\begin{equation}\label{eq:cT}
		c_{ft}' = \frac{T}{\frac{1}{2}\rho U_f^2 A_{plan}},
		\end{equation}
		where $\rho$ is the air density, $T$ is the thrust from each turbine, the planform area associated with one turbine is $A_{plan} = s_x s_y D^2$, and $s_x, s_y$ are the array spacings (normalized by the turbine diameter $D$) in the two horizontal directions.} {The prime in $c_{ft}'$ emphasizes the fact that a local fluid velocity (rather than freestream) is used (consistently with notation, for example, in Ref.~\cite{Meyers_Meneveau_WE_2012}).}
	
	In accordance with studies of vegetation canopies \cite{Huthoff_etal_WRR_2007, Chen_etal_WRR_2013}, we assume that the combination of Reynolds and dispersive stresses can be parameterized as 
	\begin{equation}
	- \left( \langle\overline{u'w'}\rangle + \langle\overline{u}''\overline{w}''\rangle \right)_{z=h_f} =  C_M  (U_b - U_f)^2
	\end{equation}
	where $C_M$ is a momentum-exchange coefficient, which is expected to vary with the geometry of the array, the design of the turbines, Reynolds number, and ambient stratification (as encoded by an appropriate Froude number). A brief  discussion of this parameterization, including possible appropriate values for $C_M$ (in a wind farm and in other canopy flows) is provided below in Sec.~\ref{sec:getCM}. We will also develop a simple heuristic model of momentum exchange, for the specific case in which the farm interface consists of a classical mixing layer. 
	
	Neglecting viscous stresses, the integral form of the momentum equation inside the wind farm is therefore
	\begin{equation}\label{eq:momF}
	\frac{c_{ft}'}{2} U_f^2 = C_M \left(U_b - U_f \right)^2.
	\end{equation}
	Following a derivation almost identical to the one described in the previous section, the flow in the overlaying boundary layer is described by the following integral forms of the mass and momentum equations (assuming $U_o>U_b>U_f$)
	\begin{eqnarray}
	\frac{d}{d x} \left( h_b U_b \right) &=& E (U_o-U_b), \label{eq:massB}\\
	\frac{d}{dx} (h_b U_b^2) &=& E \, U_o (U_o-U_b) -C_M(U_b-U_f)^2. \label{eq:momB}
	\end{eqnarray}
	To summarize, in this fully-developed regime, the horizontal momentum that is drained by the wind farm (and ultimately converted into electrical power) is replenished by vertical turbulent exchange processes with the boundary layer above the turbine array, which therefore take place at the `farm interface', as sketched in Fig.~\ref{fig:FarmSchematic}.

	Since $U_f$ does not change with downstream distance, (\ref{eq:momF}) implies that $U_b$ must also be independent of $x$. As the boundary layer grows, outer fluid is entrained at the upper edge of the boundary layer, and the vertical extent $h_b$ of the boundary layer is expected to increase. }
To get an explicit expression for $U_f$ as a function of $(c_{ft}', E, C_M)$, we first take $U_b$ out of the derivatives on the left-hand sides. Then multiplying (\ref{eq:massB}) by $U_b$ and subtracting (\ref{eq:momB}) to eliminate $d{h_b}/dx$, we obtain
\begin{equation}\label{eq:step1}
C_M(U_b-U_f)^2 = E(U_o-U_b)^2.
\end{equation}
Since we assumed that $U_o>U_b>U_f$, taking the square root and rearranging we get a linear relation between $U_b$ and $U_f$
\begin{equation}\label{eq:step2}
U_b = \frac{(C_M/E)^\frac{1}{2} U_f +U_0}{1+(C_M/E)^\frac{1}{2}}.
\end{equation}
Similarly, taking the square root of (\ref{eq:momF}) yields $U_b = U_f \{ 1+[c_{ft}'/(2C_M)]^{1/2} \}$. Subtracting this from (\ref{eq:step2}) to eliminate $U_b$, and rearranging for $U_f$, we obtain
\begin{eqnarray}\displaystyle
U_f &=& \frac{U_o}{\left(C_M^{-\frac{1}{2}}+E^{-\frac{1}{2}} \right)\left(\frac{c_{ft}'}{2}\right)^\frac{1}{2} +1}, \label{eq:UF} \\
U_b &=& U_f \left[1+ \left(\frac{c_{ft}'}{2C_M}\right)^\frac{1}{2} \right], \label{eq:UB}\\
\frac{dh_b}{dx} &=& E \frac{U_o-U_b}{U_b}. \label{eq:hB}
\end{eqnarray}
Note that, since $U_o$ and $U_b$ are both independent of $x$, (\ref{eq:hB}) implies that the boundary layer height increases linearly. As a first check of our theory, we note that this linear growth, for $h_b$, is consistent with experimental results for vegetation canopies (see, for example, Fig.~6 of Ref.~\cite{Morse_etal_bLM_2002}). 

The power density (corresponding to the power extracted per unit land area) is given by $(c_{ft}'/2) \rho U_f^3$, to a first approximation. We define a nondimensional power density coefficient, for the whole farm, by normalizing the power density (say, $P_{farm}/A_{farm}$) using the energy flux in the outer region, such that
\begin{equation}\label{eq:CfarmDef}
c_{fp} = \frac{P_{farm}/A_{farm} }{\frac{1}{2}\rho U_o^3  }=  c_{ft}' \left(\frac{U_f}{U_o} \right)^3 
\end{equation}
and, using (\ref{eq:UF}), we obtain the following concise expression:
\begin{equation}\label{eq:Cfarm}
c_{fp} = \frac{c_{ft}'}{\left[\left(C_M^{-\frac{1}{2}}+E^{-\frac{1}{2}} \right)\left(\frac{c_{ft}'}{2}\right)^\frac{1}{2} +1 \right]^3}.
\end{equation}

\paolo{
Finally, we note that while the theory presented so far neglects the effect of bottom friction (as may be reasonable to do for offshore wind farms \cite{Meneveau_JOT_2012}), it is not difficult to introduce ground dissipation effects, by adding a force, within the canopy, equal to $(c_d'/2)\rho U_f^2$, where $c_d'$ is a bottom drag coefficient. Then it is sufficient to replace $c_{ft}'$ with $c_{ft}'+c_d'$ in equations (\ref{eq:UF}) and (\ref{eq:UB}) for $U_f$ and $U_b$, and so
\begin{eqnarray}\displaystyle
U_f &=& \frac{U_o}{\left(C_M^{-\frac{1}{2}}+E^{-\frac{1}{2}} \right)\left(\frac{c_{ft}'+c_d'}{2}\right)^\frac{1}{2} +1}, \label{eq:UFdiss} \\
U_b &=& U_f \left[1+ \left(\frac{c_{ft}'+c_d'}{2C_M}\right)^\frac{1}{2} \right], \label{eq:UBdiss}
\end{eqnarray}
whereas $c_{fp}$ remains equal to $c_{ft}' (U_f/U_o)^3$ (where $U_f$ is of course now given by (\ref{eq:UFdiss})), leading to
\begin{equation}\label{eq:CfarmDiss}
c_{fp} = \frac{c_{ft}'}{\left[\left(C_M^{-\frac{1}{2}}+E^{-\frac{1}{2}} \right)\left(\frac{c_{ft}'+c_d'}{2}\right)^\frac{1}{2} +1 \right]^3}.
\end{equation}
}
To assess the accuracy of our expression for $c_{fp}$, we need to first estimate $E$, $C_M$, $c_{ft}'$ and $c_{d}'$ for existing wind farms. For the entrainment coefficient $E$, we note that a wide range of measurements in turbulent flows have found that, at very large Reynolds numbers, the entrainment coefficient is of the order \paolo{$E \sim 0.1$}  (see for example the large dataset compiled in Ref.~\cite{Cenedese_Adduce_JPO_2010} for ocean overflows, which is reproduced in Sec.~\ref{sec:stab}). The calculations needed to estimate $C_M$ and to evaluate $(c_{ft}',c_{fp})$ from field data are described in the next two subsections.

\subsection{{ Estimating $C_M$ for an interface comprising a classical mixing layer}}\label{sec:getCM}
{As noted earlier, parameterizations of the form  $C_M(U_b-U_f)^2$, for the vertical momentum flux, have been widely employed in the context of modeling canopy flows \cite{Chen_etal_WRR_2013}. One way to argue that this is a relevant scaling, specifically in a turbulent shear flow, is to use the classic closure ${-\overline{u'w'} \propto |\partial \overline{u}/\partial z| (\partial\overline{u}/\partial z)}$ and note that the velocity scale for the gradient is $(U_b-U_f)$, such that
\begin{equation}
	 -\overline{u'w'} \propto |U_b-U_f|(U_b-U_f).
\end{equation}%
To estimate $C_M$, we need a more specific physical model for the flow dynamics at the farm interface (which separates the farm and the boundary layer). Here we present a simple model, which may also be applicable to the study of other canopy flows. Recent field measurements suggest that the farm interface comprises vortical structures which closely resemble a mixing layer \cite{Hong_etal_NC_2014}. {We therefore assume a classical mixing layer structure, with thickness much smaller than $h_b$ or $h_f$, such that the mixing layer dynamics might be considered separately from the overall wind farm flow, and subsumed in a particular choice of $C_M$.} We expect that, after each turbine, the mixing layer grows downstream, entraining fluid from the farm and from the boundary layer, and mixing together the entrained fluid from these two regions, as sketched in Fig.~\ref{fig:mixingLayer}(a). The characteristic mixing layer velocity is taken as $U_{mix} = (U_b+U_f)/2$, in accordance with classical mixing-layer models \cite{Pope_CUP_2000}. As the mixing layer grows, {its thickness $\delta_{mix}$} and momentum flow $M_{mix}$ increase with distance downstream, as the layer entrains fluid from each side (see Fig.~\ref{fig:mixingLayer}):
\begin{equation}\label{eq:Mmix}
{\frac{d}{dx} (\rho\delta_{mix} U_{mix}^2) =} \frac{d M_{mix}}{dx} = E\rho(U_b-U_{mix})U_b + E\rho(U_{mix}-U_f)U_f 
= \frac{E}{2}\rho(U_b^2-U_f^2).
\end{equation}

\begin{figure}[t]
\centering
	\includegraphics[width=0.9\textwidth]{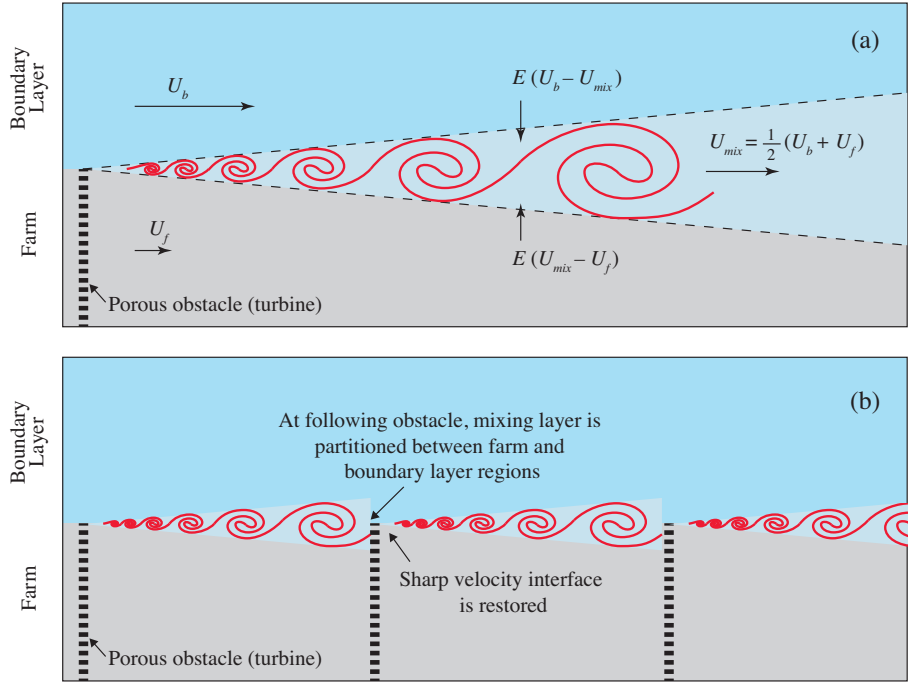}
    \vspace{10pt}
	\caption{($a$) Sketch of a spatially developing mixing layer, separating regions moving at velocities $U_b$ and $U_f$. The characteristic velocity of the mixing layer is $U_{mix} = \frac{1}{2} (U_b+U_f)$. Fluid from the top and bottom regions is entrained with velocities of magnitude $E(U_b-U_{mix})$ and $E(U_{mix}-U_f)$, respectively. {($b$) Schematic diagram of the flow past a sequence of porous obstacles. As each shear layer impinges on the following obstacle, its momentum is partitioned between the farm and boundary layer regions, and a new mixing layer starts.}}\label{fig:mixingLayer}
\end{figure}

Note that the mixing layer overlaps with both the farm and boundary layer regions. We assume that, to a first approximation, the mixing layer protrudes by the same distance into the two regions, {and, when encountering the following turbine, is partitioned equally between the two domains, as sketched in Fig.~\ref{fig:mixingLayer}(b), thereby preventing the mixing layer from growing continuously along the length of the farm.} The net momentum flux into the farm can therefore be expressed by the loss to the mixing layer, together with a gain corresponding to half the overall momentum flux into the mixing layer, i.e.
\begin{equation}\label{eq:CMder}
\mathrm{Momentum\; flux\; into\; farm}= -E\rho(U_{mix}-U_f)U_f+\frac{1}{2}\frac{d M_{mix}}{dx} = \frac{E}{4}\rho(U_b-U_f)^2.
\end{equation}

This provides a simple heuristic argument suggesting that the momentum flux from the boundary layer to the farm is proportional to $(U_b - U_f)^2$. 
Incidentally, the functional dependence that we obtained from the entrainment hypothesis can also be shown to be consistent with the well-known experimental result \cite{Brown_Roshko_JFM_1974}, for mixing layers at high Reynolds numbers, that the mixing layer thickness (say, $\delta_{mix}$) grows in proportion to \mbox{$(U_b - U_f)/(U_b+U_f)$}. To see this, since $dU_{mix}/dx=0$, (\ref{eq:Mmix}) implies
\begin{equation}
\frac{d \delta_{mix}}{dx} U_{mix}^2 = \frac{E}{2} (U_b^2-U_f^2),
\end{equation}
such that expanding {$U_{mix} = \frac{1}{2} (U_b+U_f)$} and rearranging gives
\begin{equation}
\frac{d \delta_{mix}}{dx} = 2E \frac{U_b-U_f}{U_b+U_f},
\end{equation}
consistently with the classic experimental results of Ref.~\cite{Brown_Roshko_JFM_1974}. {Heuristically, one may expect $C_M$ to be independent of $c_{ft}'$, to a first approximation, since the immediate effect of increasing $c_{ft}'$ (without otherwise changing the geometry, or drastically changing the Reynolds number) is to reduce $U_f$, and therefore increase the velocity difference across the interface. However, for mixing layers without stratification, experimental evidence indicates that $E$ is largely independent of the velocity difference \cite{Brown_Roshko_JFM_1974}.}

Furthermore, comparing (\ref{eq:CMder}) to the momentum flux parameterization $C_M \rho (U_b-U_f)^2$ used in Sec.~\ref{sec:model}, we can deduce that $C_M = E/4$ in contemporary wind farms, {provided the farm interface resembles a mixing layer}. Since $E\sim 0.1$, we note that the value of $C_M$ from our estimate is also consistent with empirical values for vegetation canopies \cite{Chen_etal_WRR_2013}. {Note, however, that canopies comprising obstacles with more complex geometries and layouts may have different values of $C_M$. Indeed, we believe that a key goal in designing new wind farms should be that of maximizing $C_M$, as discussed later in Sec.~\ref{sec:disc}.}

\subsection{Using field measurements to calculate $c_{ft}'$ and $c_{fp}$ for large wind farms}\label{sec:getData}
Following classical wind turbine theory, the thrust from one turbine is:
\begin{equation}\label{eq:T}
T = C_t \frac{1}{2}\rho U_{\infty}^2  A_{rotor},
\end{equation}
where $C_t$ is the turbine’s thrust coefficient and $A_{rotor} = \pi D^2/4$. Note that the conventional definition of $C_t$ makes use of the freestream velocity upstream of the turbine (at rotor height), which we denote as $U_{\infty}$. As noted in Ref.~\cite{Calaf_etal_PF_2010}, to relate this to the velocity at the rotor, one can use classical actuator disk theory:
\begin{equation}\label{eq:Urotor}
U_{rotor} = \frac{U_{\infty}}{2}\left(1+\sqrt{1-C_t} \right).
\end{equation}
Consistently with wind farm models based on the effective-roughness concept \cite{Calaf_etal_PF_2010,Frandsen_etal_WE_2006}, we assume that $U_{rotor} \simeq U_f$. We can combine (\ref{eq:T}) and (\ref{eq:Urotor}) with (\ref{eq:cT}) to obtain
\begin{equation}\label{eq:cTdata}
c_{ft}' = \frac{C_t \pi}{s_x s_y \left(1+\sqrt{1-C_t} \right)^2 },
\end{equation}
which enables the estimation of $c_{ft}'$ from field measurements.

Now consider $c_{fp}$, which we defined in Sec.~\ref{sec:model} as
\begin{equation}
c_{fp} = \frac{P}{\frac{1}{2}\rho U_o^3 A_{plan}},
\end{equation}
where $P$ is the power output from one turbine. In studies of turbine arrays, $P$ is typically reported as a fraction of $P_1$, which is the power output of the first turbine in the row. This makes it convenient to write $P = (P/P_1) P_1$. Using classical turbine theory, $P_1$ can be written: 
\begin{equation}
P_1 = C_p \frac{1}{2}\rho U_{\infty}^3 A_{rotor},
\end{equation}
where $C_p$ is the power coefficient. Therefore, $c_{fp}$ is given by:
\begin{equation}
c_{fp} = \left(\frac{P}{P_1}\right) \frac{C_p\pi}{4s_x s_y} \left(\frac{U_{\infty}}{U_o} \right)^3.
\end{equation}

{
	Unfortunately, $U_o$ (the velocity far above the wind farm) is not easily measured in the field. Here we use the classical power-law approximation $U(z) = U_{\infty} (z/h_{hub})^\alpha$. To decrease sensitivity to the choice of $z$, we take $U_o$ as the average between $z = h_f$ and $z = 2h_f$, such that $U_o/U_{\infty} = (2^{\alpha+1}-1)/(\alpha+1)$.
}
We first calculate $c_{fp}$ from the field measurements of Refs.~\cite{Barthelmie_etal_JAOT_2010}, \cite{Dahlberg_Thor_EOWE_2009}. Since these consist of long-time averages, and the corresponding locations exhibit predominantly neutral conditions \cite{Barthelmie_etal_JAOT_2010, Nilsson_etal_WE_2015}, we neglect the effect of ambient stratification at this stage, and set $\alpha \simeq 0.12$ (see Ref.~\cite{Nilsson_etal_WE_2015}). The key parameters used in calculating $c_{ft}'$ and $c_{fp}$ are summarized in Table~\ref{table:data}; the resulting values are displayed by the symbols in Fig.~\ref{fig:Cfarm}. 

{
	Note that there is appreciable uncertainty in estimating $U_o/U_{\infty}$. Table~\ref{table:dataUnc}  provides a breakdown of the main sources of uncertainty in the calculation required to obtain $c_{fp}$. For the uncertainty $\delta(P/P_1)$ we use available published data, whereas variations of $20\%$ are assumed for the wind exponent and upper bound of integration of the wind power-law to obtain $U_o$. The overall uncertainty is calculated by the root-sum-square method, and is shown by the error bars in Fig.~\ref{fig:Cfarm}, which combine the uncertainty from the original measurement together with the uncertainty that is introduced through the calculation of $c_{fp}$.}

\begin{sidewaystable}[p!]
\small
\centering
	\begin{tabular}{lcccccccccccc}
		Source  & \quad$  P/P_1$\quad & \quad$C_p $ \quad & \quad$s_x$ \quad &\quad $s_y$\quad  &\quad $h_f/H$\quad  & \quad $\displaystyle\frac{U_{o,\infty} }{ U_{\infty} }$ \;& $\displaystyle\frac{U_{o} }{ U_{o,\infty}}$ \; & $c_{ft}'$ & $\ c_{fp} \times 10^3$ \;& Ref. & Case & Symbol    \\ \hline
		Horns Rev & 0.63 & 0.44 & 7    & 7   & N/A & 1.11  &  1  & 0.0249 & $3.24 $ & \cite{Barthelmie_etal_JAOT_2010} & ``Aligned'' & \textcolor{blue}{$\bullet$} \\
		Nysted    & 0.61 & 0.45 & 10.3 & 5.8 &N/A& 1.11  &  1  & 0.0233 & $2.63 $ & \cite{Barthelmie_etal_JAOT_2010} & ``Aligned'' &\textcolor{red}{$\bullet$}\\
		Lillgrund & 0.30 & 0.42 & 4.3  & 3.3 &N/A& 1.12  & 1   & 0.0863 & $5.03 $ & \cite{Dahlberg_Thor_Report_2009} & ``Aligned'' &\textcolor{green}{$\bullet$}\\ \hline
		%
		Experiment & 0.558 & 0.328 & 6    & 3  &0.15 & 1.11  & 1.10  & 0.0582$^\dagger$ & 4.34 & \cite{Hamilton_Cal_PF_2015}& ``Uniform'' & \textcolor{red}{-}\\
		Experiment & 0.595 & 0.316 & 6    & 3  &0.15 & 1.11  & 1.10  & 0.0582$^\dagger$ & 4.46 & \cite{Hamilton_Cal_PF_2015} & ``Row-by-row'' &\textcolor{blue}{+}\\
		Experiment & 0.664 & 0.305 & 6    & 3  &0.15 & 1.11  & 1.10  & 0.0582$^\dagger$ & 4.81 & \cite{Hamilton_Cal_PF_2015} & ``Col.-by-col.'' &\textcolor{black}{*}\\
		Experiment & 0.623 & 0.305 & 6    & 3  &0.15 & 1.11  & 1.10  & 0.0582$^\dagger$ & 4.51 & \cite{Hamilton_Cal_PF_2015} & ``Checkerboard'' &\textcolor{green}{$\times$}\\
		\hline
		
		
		Full-farm LES & 0.550 & 0.4$^\ddagger$ & 7.85  & 3.49 & 0.075& 1.11$^\S$ & 1.07&0.0382 & 3.82 & \cite{Stevens_etal_JRSE_2014} & ``Aligned'' &$\triangleright$\\
		Full-farm LES & 0.569 & 0.4$^\ddagger$ & 7.85  & 3.49 & 0.075& 1.11$^\S$ & 1.07&0.0382 & 3.95 & \cite{Stevens_etal_JRSE_2014} & ``Staggered''&$\diamond$\\
		Full-farm LES & 0.619 & 0.4$^\ddagger$ & 7.85  & 5.23 & 0.075& 1.11$^\S$ & 1.06&0.0255 & 2.91 & \cite{Stevens_etal_JRSE_2014} & ``Aligned''&$\triangleright$\\
		Full-farm LES & 0.651 & 0.4$^\ddagger$ & 7.85  & 5.23 & 0.075& 1.11$^\S$ & 1.06&0.0255 & 3.07 & \cite{Stevens_etal_JRSE_2014} & ``Staggered''&$\diamond$\\
		
		Full-farm LES & 0.549 & 0.445 & 7  & 7 &0.105& 1.13 & 1.07&0.0263 & 2.20 & \cite{Wu_Porte-Agel_RE_2015} & ``$\pm 1^\circ$'' &$\triangle$\\
		
		Full-farm LES & 0.655 & 0.4$^\ddagger$ & 7.85  & 5.24 &0.075& 1.11$^\S$ & 1.05&0.0255 & 3.13 & \cite{Stevens_WE_2015} & ``A3, aligned'' &$\square$\\
		Full-farm LES & 0.580 & 0.4$^\ddagger$ & 5.24  & 5.24 &0.075& 1.11$^\S$ & 1.06&0.0381 & 4.07 & \cite{Stevens_WE_2015} & ``C4, aligned'' &$\square$\\
		Full-farm LES & 0.462 & 0.4$^\ddagger$ & 5.24  & 3.49 &0.075& 1.11$^\S$ & 1.07&0.0573 & 4.74 & \cite{Stevens_WE_2015} &  ``D3, aligned'' &$\square$\\
		
		Full-farm LES & 0.660 & 0.4$^\ddagger$ & 7.85 & 5.24 &0.075& 1.11$^\S$ & 1.05&0.0255 & 3.16 & \cite{Stevens_etal_WE_2015} & ``A3, staggered'' & $\triangledown$ \\
		Full-farm LES & 0.563 & 0.4$^\ddagger$ & 5.24 & 5.24 &0.075& 1.11$^\S$ & 1.06&0.0382 & 3.96 & \cite{Stevens_etal_WE_2015} & ``C4, staggered'' & $\triangledown$ \\
		Full-farm LES & 0.542 & 0.4$^\ddagger$ & 3.49 & 7.85 &0.075& 1.11$^\S$ & 1.06&0.0382 & 3.81 & \cite{Stevens_etal_WE_2015} & ``F3, staggered'' & $\triangledown$ \\
		Full-farm LES & 0.453 & 0.4$^\ddagger$ & 3.49 & 5.24 &0.075& 1.11$^\S$ & 1.07&0.0573 & 4.66 & \cite{Stevens_etal_WE_2015} & ``G3, staggered'' & $\triangledown$ \\
		Full-farm LES & 0.783 & 0.4$^\ddagger$ & 7.85 & 7.85 &0.075& 1.11$^\S$ & 1.05&0.0170 & 2.55 & \cite{Stevens_etal_WE_2015} & ``H3, staggered'' & $\triangledown$ \\
		
		\hline
	\end{tabular}
    \vspace{10pt}
	\caption{Wind farm data used in the preparation of Fig.~\ref{fig:Cfarm}. Values of $P/P_1$ for Horns Rev and Nysted correspond to case `ER', from Fig.~5 of Ref.~\cite{Barthelmie_etal_JAOT_2010}. Lillgrund data for $P/P_1$ is adapted from Fig.~30 of Ref.~\cite{Dahlberg_Thor_Report_2009}. {Power data for the last row is used in each case. $\dagger$ assumes $C_t\simeq 0.75$, $\ddagger$ assumes $C_p\simeq 0.4$. $\S$ uses $U_o/U_{\infty}$ from Ref.~\cite{Stevens:2014ia}, where the precursor method is discussed.}}\label{table:data}
\end{sidewaystable}

\begin{table}[t!]
    \small
	\begin{tabular}{lcccccccc}
		Site  & $ \left[ \frac{\delta (P/P_1)}{P/P_1} \right.$ & $\left. \frac{\delta c_{fp}}{c_{fp}} \right]$   & $\left[\frac{\delta\alpha}{\alpha}\right.$   & $\left. \frac{\delta c_{fp}}{c_{fp}} \right]$  & $ \left[ \frac{\delta (z_o/h_f)}{z_o/h_f} \right.$ & $\left. \frac{\delta c_{fp}}{c_{fp}} \right]$     & Overall $ \frac{\delta c_{fp}}{c_{fp}} $ & Ref.     \\ \hline
		Horns Rev & [0.29 & 0.29] & [0.2 & 0.03] &[0.2 & 0.48] & 0.56 & \cite{Barthelmie_etal_JAOT_2010} \\
		Nysted    & [0.19 & 0.19] & [0.2 & 0.03] &[0.2 & 0.48] & 0.51 & \cite{Barthelmie_etal_JAOT_2010} \\
		Lillgrund & [0.085& 0.085]& [0.2 & 0.03] &[0.2 & 0.48] & 0.49 & \cite{Dahlberg_Thor_Report_2009} \\ \hline
	\end{tabular}
    \vspace{10pt}
	\caption{Main sources of uncertainty, and their impact on $c_{fp}$, for field measurements. The variation in $P/P_1$ is from the published data for these turbine sites \cite{Barthelmie_etal_JAOT_2010,Dahlberg_Thor_Report_2009}. For the wind exponent $\alpha$ and height $z_{o}/h_f$ for estimating $U_o$, an uncertainty of $20\%$ relative to reference values is assumed. Overall uncertainties in $c_{fp}$ are estimated by the root-sum-square method {\cite{Moffat:1988vu}}.}\label{table:dataUnc}
\end{table}

\subsection{\paolo{Estimating a blockage correction for experiments and full-farm LES}}
%
%
\paolo{For wind tunnel measurements, even in an empty test section, boundary layers effectively reduce the cross sectional area available to the flow, causing the outer flow velocity to gradually increase with downstream distance. This effect is of course stronger for arrays of obstacles, since their drag causes the flow to slow down locally, thereby forcing a substantial amount of fluid out of the array and into the outer flow, such that $U_o$ is increased. Consistently with classic work on wings and bluff-bodies~\cite{Maskell:1963vy}, the resulting blockage is greater than the value that would be obtained by just considering the contraction associated with the cross-sectional area of the objects. This is true in experiments~\cite{Hamilton_Cal_PF_2015} as well as simulations that employ a no-through-flow boundary condition at the top of the domain~\cite{Stevens_WE_2015,Stevens_etal_JRSE_2014,Stevens_etal_WE_2015,Wu_Porte-Agel_RE_2015}. The outer flow velocity $U_o$ is typically not reported, but the inflow profile is described. This must be related to the outer flow velocity $U_o$ above the wind farm, at a location where the farm flow is fully developed. While $U_o$ increases with downstream distance, we assume that $U_o$ varies slowly, by comparison to the distance over which turbulent fluxes equilibrate, so that the steady model from Sec.~\ref{sec:model} still applies.}

\paolo{
In the spirit of classic blockage models \cite{Maskell:1963vy}, and consistently with the model in Sec.~\ref{sec:model}, we use a layer-wise approach, illustrated in Fig.~\ref{fig:blockage}. Continuity implies
\begin{equation}\label{eq:blockage1}
U_{\infty} h_f + U_{o,\infty} (H-h_f)  = U_f h_f + U_b h_b + U_o (H-h_f-h_b)
\end{equation}
where $H$ is the height of the test section. Equation~(\ref{eq:blockage1}) is written as
\begin{equation}
U_{o,\infty} \left[ \frac{U_{\infty}}{U_{o,\infty}} h_f +  (H-h_f)\right]  = U_{o} \left[ \frac{U_f}{U_o} h_f + \frac{U_b}{U_o} h_b + (H-h_f-h_b)\right],
\end{equation}
such that, dividing both sides by $h_f$ and solving for the blockage correction factor $U_o/U_{o,\infty}$ 
\begin{equation}\label{eq:blockageCorr}
\frac{U_{o,\infty}}{U_{o}} =\frac{\displaystyle \frac{U_{\infty}}{U_{o,\infty}} +  \left(\frac{H}{h_f}-1\right)}{\displaystyle \frac{U_f}{U_o} + \frac{U_b}{U_o} \frac{h_b}{h_f} + \left(\frac{H}{h_f}-1-\frac{h_b}{h_f}\right)}.
\end{equation}
To evaluate~(\ref{eq:blockageCorr}), we use (\ref{eq:UFdiss}-\ref{eq:UBdiss}) for $U_f/U_o,U_b/U_o$, and approximate $h_b/h_f \simeq (dh_b/dx)(x/h_f)$, where $dh_b/dx$ is found from (\ref{eq:hB}) and $x/h_f$ depends on the dataset. The blockage correction is reported in Table~\ref{table:data}. Note that while a value of $E$ must be assumed, the blockage correction has only very weak sensitivity to the entrainment coefficient, since changing $E$, say, from $0.1$ to $0.16$ yields a variation of less than 1\% in~(\ref{eq:blockageCorr}).
}

 \begin{figure}[t]
 \centering
 	\includegraphics[width = 0.8\textwidth]{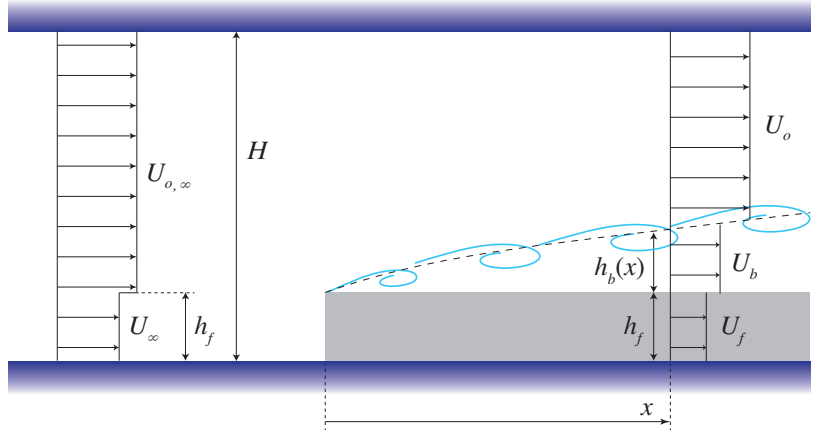}
    \vspace{10pt}
 	\caption{\paolo{Schematic illustration of blockage effects on the outer velocity $U_o$, in wind tunnel measurements and complete-farm simulations.}
 	} \label{fig:blockage}
 \end{figure}


\subsection{Estimating the bottom drag coefficient $c_d'$}\label{sec:cd}
\paolo{From (\ref{eq:T},\ref{eq:vertAverage}), the stress due to the combined effect of the wind farm and of bottom drag is
\begin{equation}\label{eq:draglaw}
\tau = \left(c'_{ft}+ c'_d\right)\frac{\rho}{2} \left(\frac{1}{h_f} \int_0^{h_f} \langle \overline{u}\rangle \,dz \right)^2 = \rho \left(u^* \right)^2.
\end{equation}
Consider the flow ahead of the wind farm, where $c'_{ft}=0$. For $z<h_f$, the log-law is a reasonable approximation \cite{Hamilton_Cal_PF_2015}, such that
\begin{equation}\label{eq:loglaw}
\langle \overline{u} \rangle = \frac{u^*}{\kappa} \ln \frac{z}{z_0},
\end{equation}
where $\kappa$ is the von K\'arm\'an constant (approximately equal to 0.4) and $z_0$ is the equivalent roughness height for the bottom surface, which is assumed known. Substituting~(\ref{eq:loglaw}) into~(\ref{eq:draglaw}), integrating, and solving for $c_d'$ yields
\begin{equation}
c_d' = \frac{2 \kappa^2}{\left(1+\ln \frac{z_0}{h_f} \right)^2}.
\end{equation}
Values of $c'_d$ do not vary significantly in experiments and simulations. For example, \cite{Hamilton_Cal_PF_2015} report $z_0/h_f \simeq 5.56\times 10^{-4}$, implying $c_d'\ \simeq 0.0076$, whereas \cite{Stevens:2014ia} have $z_0/h_f \simeq 9.77\times 10^{-4}$, yielding  $c_d'\ \simeq 0.0091$. For simplicity, in our theory, we take $c_d' = 0.008$.
}

\subsection{Comparing the two-interface model with  measurements}
\paolo{
In order to perform a comparison between our theory and measurements, we still need to select a value for the entrainment coefficient $E$. For large-scale turbulent flows (such as gravity currents, jets and plumes) the entrainment coefficient is typically of the order $E \sim 0.1$ (e.g.~Ref.~\cite{Cenedese_Adduce_JPO_2010}). There appear to be no precise published values of $E$ for turbulent boundary layers. To determine a value of $E$, we use published data for the growth of a boundary layer over wind turbine arrays. Since boundary layer thickness is not measured in full-scale wind farms (to the best of our knowledge), we compare our solution to the wind tunnel results of Ref.~\cite{Markfort:2012ec} for aligned and staggered wind turbines. In order to show an example at larger $c'_{ft}+c'_d$ also, we include the boundary layer for a flume experiment with a model canopy comprising vertical cylinders, from Ref.~\cite{Chen_etal_WRR_2013}. To express the boundary layer growth in terms of a standard quantity for boundary layers, consider the definition of the displacement thickness $\delta^*$~\cite{Schlichting_McGraw-Hill_1979}, and use equation~(\ref{eq:def1}) 
\begin{equation}
\delta^* = \int_0^H \left(1 - \frac{\langle \overline{u} \rangle}{U_o} \right) \, dz = \left(1 - \frac{U_b}{U_o} \right) \, h_b.
\end{equation}
Differentiate by $x$, using the fact that $d U_b/dx=0$ in fully-developed flow
\begin{equation}
\frac{d\delta^*}{dx} = \left(1 - \frac{U_b}{U_o} \right) \,\frac{d h_b}{dx}, 
\end{equation}
where $U_b/U_o$ and $dh_b/dx$ are found from~(\ref{eq:UBdiss},\ref{eq:hB}). Figure~\ref{fig:dhBdx} shows that the two-interface theory with $E = 0.16$ and $C_M = E/4 = 0.04$ is in agreement with experimental data.}

\begin{figure}
		\centering
		\includegraphics[width=0.95\textwidth]{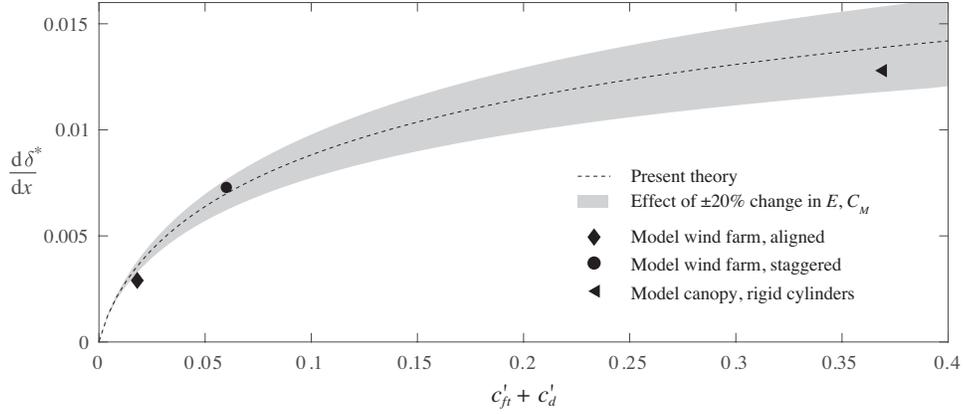}
		\caption{\paolo{Displacement thickness growth rate $d\delta^*/dx$ as a function of the combined planform-averaged thrust and drag coefficients, that is $c'_{ft}+c'_d$. The dashed line corresponds to our model, assuming $E=0.16$ and $C_M=0.04$. Symbols are laboratory experiments for model wind farms \cite{Markfort:2012ec} as well as for model canopies \cite{Chen_etal_WRR_2013}. } }
		\label{fig:dhBdx}
\end{figure}

We now use equation (\ref{eq:CfarmDiss}) and the values $E = 0.16$ and $C_M = E/4 = 0.04$ to {obtain an approximate value} of $c_{fp}$, as a function of $c_{ft}'$, for existing wind farms, and compare it to the data summarized in Table~\ref{table:data}. The result is shown by the dashed line in Fig.~\ref{fig:Cfarm}($a$), where we take $c_d' = 0.008$, following the analysis in Sec.~\ref{sec:cd}. The effect of varying $E$ and $C_M$ by $\pm 20 \%$ is shown by the shaded region. In spite of the simplicity of our model, it is in agreement with available data. 
The plot of $c_{fp}$ versus $c_{ft}'$ has a few notable features. Firstly, $c_{fp}$ is a very small number (of the order of $0.1\%$), expressing the fact that a ground-based wind farm can extract a small fraction of the power flux overhead. Secondly, as \paolo{shown in Fig.~\ref{fig:Cfarm}($b$), at larger $c_{ft}'$ the $c_{fp}$ curve reaches a maximum, which corresponds to optimizing over the possible turbine spacings and thrust settings. The value of $c_{ft}'$ that maximizes power output is found by differentiating~(\ref{eq:CfarmDiss}) and equating it to zero, yielding 
\begin{equation}\label{eq:cft_maxP}
	\left.{c_{ft}'}\right|_{\max\,c_{fp}} = 2 \left(c'_d + 2\zeta^2 \right) + 4\zeta \sqrt{\frac{3}{2}c'_d+\zeta^2}, 
\end{equation}
where $\zeta = ( C_M^{-\frac{1}{2}} + E^{-\frac{1}{2}})^{-1}$. Here, the optimal ${c_{ft}'} \simeq 0.179$, which corresponds to $c_{fp}\simeq 5.0 \times 10^{-3}$. Assuming $C_t = 8/9$ (corresponding to the Betz limit) and $s=s_x=s_y$, this would imply a turbine spacing of 3.0$D$, which is somewhat smaller than the value at the Lillgrund site (which has $s_x = 4.3$, $s_y = 3.3$), as illustrated in Fig.~\ref{fig:Cfarm}($b$). However, since the optimum in Fig.~\ref{fig:Cfarm}($b$) is very broad, there is little reason to seek such small turbine spacings. For example, the theory predicts that increasing $s$ by 50\% from the optimal value will result in a power decrease of only about 6\%.}

Note that, according to (\ref{eq:cT}), $c_{ft}'$ is inversely proportional to the square of the turbine spacing, such that, for a given land extent, an increase in $c_{ft}'$ corresponds to an increase in the number of installed turbines. This implies that increasing the number of turbines beyond a certain value can actually be detrimental to power output (with given land area), {as the lower values of $U_f$ that result do not make up for the large values of $c_{ft}'$.}

\begin{figure}[t]
	\includegraphics[width=\linewidth]{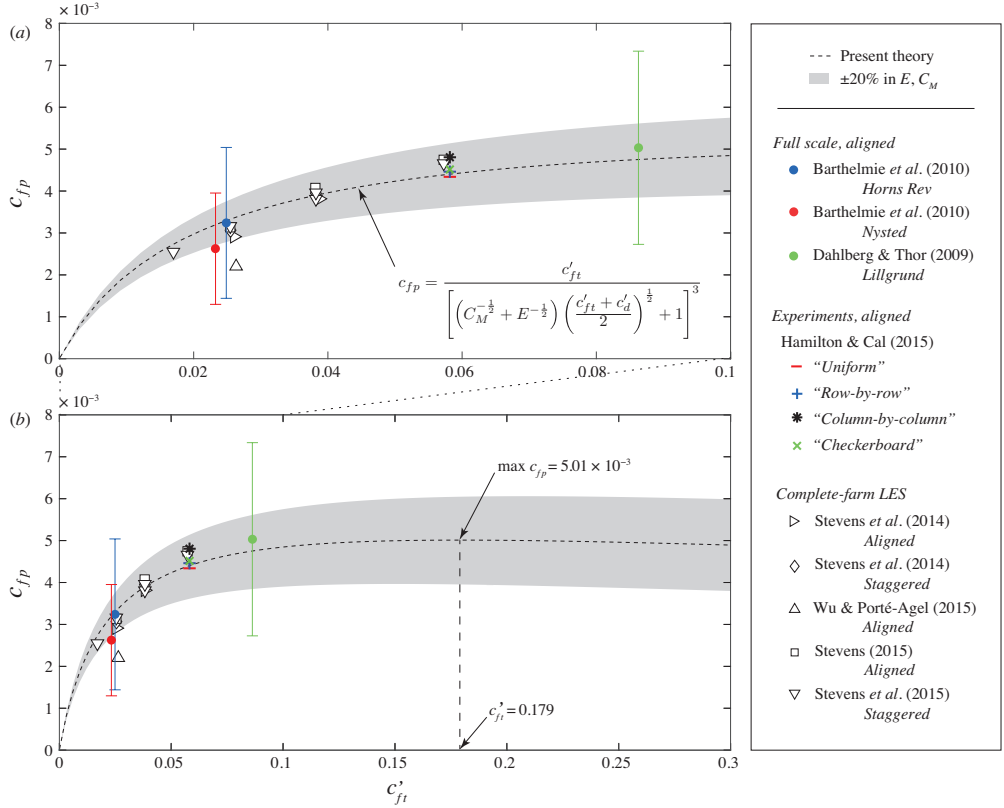}
	\caption{\paolo{Farm power coefficient versus planform-averaged thrust coefficient. Symbols correspond to sources reported in Table~\ref{table:data}; sources of uncertainty for the field measurements are described in Table~\ref{table:dataUnc}. The dashed line corresponds to our theory with $E = 0.16$, $C_M = E/4 = 0.04$ and $c_d' = 0.008$, whereas the gray region corresponds to the effect of a change of $\pm 20\%$ in $E$ and $C_M$. Panel ($a$) shows results for \mbox{$0<c_{ft}' < 0.1$}; available experimental, numerical and field data are all within this range of $c_{ft}'$. Panel ($b$) illustrates the trend predicted for larger $0<c_{ft}' < 0.3$, which exhibits a maximum for $c_{ft}' \simeq 0.179$.}
	} \label{fig:Cfarm}
\end{figure}

\section{Modeling the effect of atmospheric stability on wind farm performance}\label{sec:stab}
\subsection{Linking $E$ and $C_M$ to the Obukhov length}
Here we extend our theory to include the effect of atmospheric stability, by considering the influence of ambient conditions on the parameters $E$ and $C_M$. This approach accounts for the effect of stratification on the turbulent motions responsible for mass and momentum exchanges. {We employ a recently introduced parameterization  \cite{Cenedese_Adduce_JPO_2010}, which describes $E$ as an empirical function of the Reynolds number $Re$ and Froude number $Fr$, based on a large compilation of laboratory experiments and field observations involving ocean overflows; here we refer to this fit for $E$ using the expression $E_{ca10}(Re,Fr)$. For completeness, the dataset on which the parameterization is based is shown in Fig.~\ref{fig:parametrization}, and the fit equations are (using the same notation as Ref.~\cite{Cenedese_Adduce_JPO_2010})
	\begin{eqnarray}
	E_{ca10}(Re,Fr) &=& \frac{E_{min} +A Fr^\alpha}{1+A\, C_{inf} (Fr + Fr_0)^\alpha} \label{eq:Eclaudia}\\
	C_{inf} &=& \frac{1}{E_{max}} + \frac{B}{Re^\beta}
	\end{eqnarray}
	where $E_{min} = 4\times 10^{-5}$, $E_{max} = 1$, $A = 3.4 \times 10^{-3}$, $B = 243.52$, $\alpha = 7.18$, $\beta = 0.5$, $Fr_0 = 0.51$. In Ref.~\cite{Cenedese_Adduce_JPO_2010}, the Reynolds and Froude numbers are based on the gravity current height and on the velocity difference across the turbulent interface, and are defined as
	\begin{equation}
	Re = \frac{h\Delta U}{\nu}, \quad\quad Fr = \frac{\Delta U}{\sqrt{h g\frac{\Delta\rho}{\rho_o}}},
	\end{equation}
	where $g$ is the gravitational acceleration, $h$ is the vertical length-scale of the flow (corresponding to the vertical extent of the gravity current in ocean measurements), $\Delta U$ and $\Delta \rho$ are the velocity and density differences across the interface, and $\rho_0$ is a reference density. 
	
	From Fig.~\ref{fig:parametrization}, it is apparent that for $Re$ larger than about $10^4$, $E_{ca10}$ can exceed 0.16 for large $Fr$. However, \paolo{while there are atmospheric measurements that support $E$ as high as 0.2 \cite{Princevac_etal_JFM_2005}, published} data do not seem to support significantly larger values of $E$, as also noted in Ref.~\cite{Wells_etal_JPO_2010}. For this reason, we propose a modified version of the fit, which saturates at a value $E_{sat}$. While we could simply alter $E_{max}$ in (\ref{eq:Eclaudia}), this has the unwanted effect of altering $E_{ca10}$ for all $Re$. At fixed $Re$, we therefore simply evaluate (\ref{eq:Eclaudia}) as a function of $Fr$. Based on the value of the resulting $E_{ca10}$, our $E$ is given by
	\begin{equation}\displaystyle
	E_{par}(Fr) = \left\{ \begin{array}{ll}
	E_{ca10}(Fr) & \mbox{if $E \leq E_{cut}$};\\ [5pt] \displaystyle
	E_{cut} + \left[ \frac{\partial E_{ca10}}{\partial Fr} \right]_{cut} \frac{Fr - Fr_{cut} }{1+ \displaystyle \left[ \frac{\partial E_{ca10}}{\partial Fr} \right]_{cut}  \frac{Fr-Fr_{cut} }{E_{sat}-E_{cut}} }  & \mbox{if $E > E_{cut}$}.\end{array} \right.
	\end{equation}
	where we set $E_{cut} = 0.8 E_{sat} \paolo{= 0.128}$ and $Fr_{cut}$ is the value of $Fr$ at which $E_{ca10} = E_{cut}$, {which is found by interpolation. For $Re=10^8$, \paolo{$Fr_{cut} \simeq 1.95$}}. This provides a smooth algebraic transition between the parameterization of Ref.~\cite{Cenedese_Adduce_JPO_2010} and $E_{sat}$, as shown by the red continuous and dashed lines in Fig.~\ref{fig:parametrization}.
	
	In our flow of interest, the characteristic scale $h$ is the farm height $h_f$, and we use the classic atmospheric approximation $\Delta\rho /\rho_o = -\Delta \theta/\theta_o$,  where $\theta$ is the potential temperature. For each interface, we consider the corresponding differences in potential temperature and velocity to calculate the relevant $Fr$. Therefore we write, for each interface:
	{
		\gdef\thesubequation{\theequation \textit{a,b}}
		\begin{subeqnarray}
			\label{eq:ECM}
			E = E_{par}(Fr_o) ,\qquad C_M = \frac{1}{4} E_{par}(Fr_f), 
		\end{subeqnarray}
		where the subscripts to $Fr$ specify the interface considered (`outer' or `farm')}. } 

\begin{figure}[t]
\centering
	\includegraphics[width=0.99\textwidth]{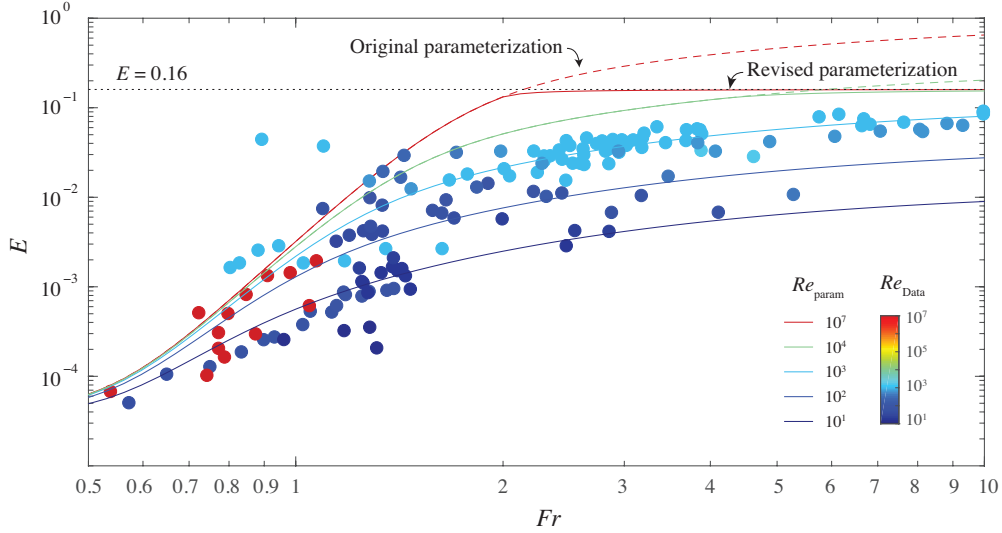}
    \vspace{10pt}
	\caption{Entrainment coefficient $E$ versus Froude number $Fr$, as $Re$ is varied, adapted from Ref.~\cite{Cenedese_Adduce_JPO_2010}, which compiles data for ocean overflows. Symbols show experimental and field data. The colored lines show the parameterization proposed in Ref.~\cite{Cenedese_Adduce_JPO_2010}, which yields $E\rightarrow 1$ at large $Re$ and finite $Fr$. Since the available data does not seem to support $E$ of order 1, here we propose a slightly revised parameterization, which caps $E$ at 0.16. The difference between the original and revised parameterizations is shown by the red dashed and continuous lines. The two parameterizations are identical for $Re$ less than about $10^4$.} \label{fig:parametrization}
\end{figure}

We assume that a constant specific heat flux $q$ is applied at the bottom of the farm. {By analogy with the momentum flux parametrization, we assume that the exchange of heat at the farm interface is proportional to $C_\theta (U_b-U_f) (\theta_b-\theta_f)$, where $C_\theta$ is an exchange coefficient analogous to $C_M$.}
Using the same subscripts as before to denote each region, the equations governing potential temperature are
\begin{eqnarray}
\frac{d}{dx}(h_b U_b \theta_b) &=& E(U_o-U_b)\theta_o - C_\theta(U_b-U_f)(\theta_b-\theta_f), \label{eq:heatB}\\
\frac{d}{dx}(h_f U_f \theta_f) &=& C_\theta (U_b-U_f)(\theta_b-\theta_f) + q/c_p. \label{eq:heatF}
\end{eqnarray}
{In the above, $c_p$ is the specific heat of air, which has been assumed constant. Consistently with our fully-developed assumption, we take the temperature in each region to be constant with downstream distance (such that the left-hand side of (\ref{eq:heatB}) is $U_b \theta_b dh_b/dx$, whereas the left-hand side of (\ref{eq:heatF}) vanishes). Then (\ref{eq:heatB}) and (\ref{eq:heatF}) are two linear equations for $\theta_f$, $\theta_b$, with constant coefficients. Using (\ref{eq:hB}) for $dh_b/dx$, these expressions can be manipulated to yield the nondimensional temperature differences relative to the outer flow:}
\begin{eqnarray}
\frac{\theta_f-\theta_o}{\theta_o} &=& \frac{q}{c_p U_o \theta_o} \left(\frac{1}{E} \frac{U_o}{U_o-U_b} + \frac{1}{C_\theta} \frac{U_o}{U_b-U_f} \right), \label{eq:thetaF}\\
\frac{\theta_b-\theta_o}{\theta_o} &=& \frac{q}{c_p U_o \theta_o} \left(\frac{1}{E} \frac{U_o}{U_o-U_b} \right). \label{eq:thetaB}
\end{eqnarray}

We now need to relate the nondimensional heat flux $q/(c_p U_o \theta_o)$ to quantities that are reported through field measurements. For example, as discussed later in the next subsection, the wind farm measurements of Ref.~\cite{Hansen_etal_WE_2011} are sorted using estimates of the Obukhov length $L$ \cite{Obukhov_BLM_1971}}
\begin{equation}\label{eq:Obukhov}
L = - \frac{c_p \theta_o \tau^{3/2}}{\kappa g \rho^{3/2} q},
\end{equation}
where $\tau$ is the shear stress in the boundary layer, and $\kappa \simeq 0.4$ is the von K\'arm\'an constant. {Physically, $L$ (given by (\ref{eq:Obukhov}) above) expresses the height above which damping from the stratification exceeds the production of turbulent kinetic energy. Since turbulent production peaks close to the ground, a smaller $L$ corresponds to an increasingly stable atmosphere \cite{Pope_CUP_2000,Obukhov_BLM_1971}.} Since here \paolo{$\tau = \frac{1}{2} \rho (c_{ft}'+c_d') U_f^2$}, we obtain
\paolo{
\begin{equation}\label{eq:q}
\frac{q}{c_p U_o \theta_o} = -\frac{h_f}{L} \frac{[(c_{ft}'+c_d')/2]^{3/2} (U_f/U_o)^3}{\kappa (g h_f / U_o^2)}.
\end{equation}
}

With given values of $c_{ft}', c_d'$ and $L/h_f$ (as well as $gh_f/U_o^2$), the resulting set of equations (\ref{eq:UF}, \ref{eq:UB}, \ref{eq:ECM}$a$,$b$, \ref{eq:thetaF}, \ref{eq:thetaB}, \ref{eq:q}) for $U_f$, $U_b$, $\theta_f$, $\theta_b$, $E$, $C_M$ and $q$ can be solved by iteration. We find that a simple relaxation method is sufficient to obtain convergence. For simplicity, we set $C_\theta = C_M$, which essentially corresponds to the assumption (common in turbulent flows {\cite{Obukhov_BLM_1971}}) that the turbulent Prandtl number is close to unity.

\begin{table}[t]
\centering
	\begin{tabular}{lccccc}
		Conditions       & $\quad P/P_1$ & $\quad L/ h_f$ interval  \quad   & $\quad U_o / U_{\infty} $ & $c_{ft}'$     & $\quad c_{fp} \times 10^3$     \\ \hline
		Very stable      & 0.30 & ${[}0.091,   0.45{]}$ & 1.57       & 0.0249 & $1.01$ \\
		Stable           & 0.61 & ${[}0.45,   1.8{]}$   & 1.47       & 0.0249 & $1.22$ \\
		Neutral/Unstable & 0.59 & ${[}1.8,   \infty{]}$      & 1.11       & 0.0249 & $3.24$ \\ \hline
	\end{tabular}
    \vspace{10pt}
	\caption{. Dependence of power output on stability conditions. Values of $P/P_1$ and corresponding intervals of $L/h_f$ are calculated from Ref.~\cite{Hansen_etal_WE_2011}. For {Horns Rev}, $h_f = 110$\,m.}\label{table:stabdata}
\end{table}

\begin{figure}[t]
	\includegraphics[width=0.99\textwidth]{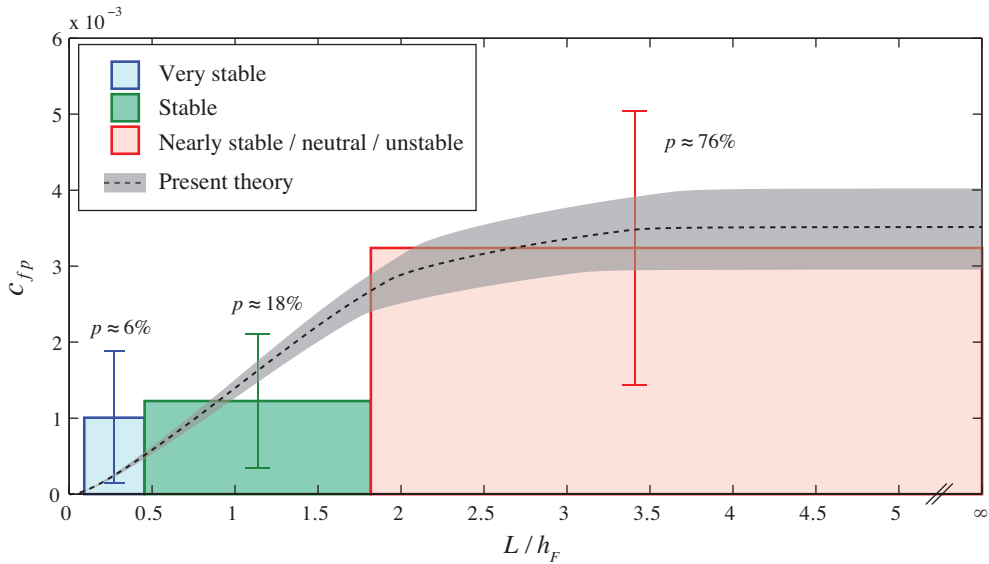}
	\caption{Effect of atmospheric stability on power production, \paolo{as expressed by} farm power coefficient in a fully-developed wind farm, as a function of Obukhov length $L$ (normalized by wind farm height $h_f$). Bars show published data, corresponding to the Horns Rev wind farm \cite{Hansen_etal_WE_2011}. The values of $p$ show the reported probabilities, for westerly winds, of each stability class. The dashed line corresponds to our theory (with $c_{ft}' = 0.0291$ matching the estimated $c_{ft}'$ for Horns Rev, and assuming $c_d' \simeq 0.008$), {whereas the gray shaded region shows the effect of varying the value of $E_{par}$, in the entrainment parameterization, by $\pm 20\%$}. Error bars as in Fig.~\ref{fig:Cfarm}.} \label{fig:stab}
\end{figure}

\subsection{Power dependence on atmospheric stability in theory and field measurements}
{ Measurements of atmospheric stability at wind turbine sites are currently quite limited. We are aware of two such published studies, namely Refs.~\cite{Hansen_etal_WE_2011, Vanderwende_Lundquist_ERL_2012}. The former provides data for the well-known Horns Rev site, whereas the latter acquired data at a wind farm in the U.S. midwest (the specific site, layout, and turbine model do not appear to have been described). No data on ambient stability is available for the Nysted and Lillgrund sites, to the best of our knowledge. To avoid complications associated with neighboring topography, here we focus on the measurements \cite{Hansen_etal_WE_2011} from the Horns Rev site.} These used estimates of the Obukhov length $L$ to categorize atmospheric conditions. The field measurements of Ref.~\cite{Hansen_etal_WE_2011} found that as the atmosphere transitioned from stable to neutral, power output increased \cite{Hansen_etal_WE_2011}. However, as the atmosphere shifted from neutral to unstable, power output did not increase further; for this reason, results were reported by binning together measurements involving nearly stable, neutral and unstable conditions \cite{Hansen_etal_WE_2011}.

Recall from Sec.~\ref{sec:getData} that extracting $c_{fp}$ from field data requires estimating $U_o$ above the wind farm. However, the value of $\alpha$ in the power-law approximation for $U_o$ depends on the Obukhov length $L$. Here we use the measurements of Ref.~\cite{Irwin_AE_1979}, for flow over relatively smooth terrain (equivalent roughness $z_0 \simeq 0.01\,$m), which report $\alpha$ as a function of atmospheric stability category. To match these values of $\alpha$ to the corresponding $L$, for {Horns Rev}, we use the field measurements of Ref.~\cite{Hansen_etal_WE_2011} (which associate $L$ intervals to stability conditions), thereby obtaining the following approximate $(\alpha, L)$ pairings: (0.53, 50\,m), (0.34, 200\,m), (0.12, 500\,m). An empirical function $\alpha(L)$ is then defined by using a piecewise cubic Hermite interpolant between these values, with constant value outside of the interval $L = [50\,\mathrm{m}, 500\,\mathrm{m}]$. The associated data are reported in Table~\ref{table:stabdata}, whereas the results are displayed in Fig.~\ref{fig:stab}, as discussed below.

Our prediction for $c_{fp}$ versus $L/h_f$ is shown by the dashed line in Fig.~\ref{fig:stab}, whereas field data \cite{Hansen_etal_WE_2011} are shown by the shaded bars. {The gray shaded region shows the sensitivity of the results to a change of $\pm 20\%$ in $E_{par}$, with given Froude numbers.} The theory captures the trend in $c_{fp}$, and is in good quantitative agreement with records for stable and neutral/unstable atmospheres. Our model seems to underpredict $c_{fp}$ for very stable conditions. This might be due to the parameterization used for $E$, which assumes that the entrainment tends to a small value as the atmosphere becomes strongly stratified, as is the case in flows without large obstacles that can maintain fluid mixing. However, in a wind farm, the presence of turbines would drive the flow towards a larger $E$. 
In addition, the turbulent Prandtl number might be deviating from unity, as the flow becomes more strongly stratified (this could make $C_\theta$ depart significantly from $C_M$). 
Interestingly, the literature on vegetation canopies indicates that traditional Monin-Obukhov theory also underestimates momentum fluxes for plants in a stable stratification, although there does not seem an accepted explanation for this underestimate; see for example Ref.~\cite{Arnqvist:2015to} and references therein.
However, for wind farms, since very stable conditions were relatively rare in the field measurements \cite{Hansen_etal_WE_2011} (corresponding to only a few percent of observations), this model limitation might not have a very significant practical impact.


\section{An ideal limit for wind farm performance}\label{sec:limit}

We now turn to the problem of finding an upper bound for the relative power density, as the design of a wind farm is varied. \paolo{The maximum of $c_{fp}$ over $c_{ft}'$ was found earlier as (\ref{eq:cft_maxP}).  Since we focus on idealized performance, assume $c_d' \ll c_{ft}'$, such that 
\begin{equation}
	\left.{c_{ft}'}\right|_{\max\,c_{fp}} = 8 \left( C_M^{-\frac{1}{2}} + E^{-\frac{1}{2}}  \right)^{-2}
\end{equation} 
and~(\ref{eq:Cfarm}) yields}
\begin{equation}
\max_{c_{ft}'}\, c_{fp} = \frac{8}{27} \left( C_M^{-\frac{1}{2}} + E^{-\frac{1}{2}}  \right)^{-2}.
\end{equation}

The parameter $E$ controls the transfer of momentum at the interface between the outer flow and the boundary layer, and is unlikely to be affected by the design of the wind farm. However, we propose that it may be possible to redesign a wind farm to increase $C_M$, by promoting exchanges between the farm and the boundary layer. To establish an upper bound on $c_{fp}$, we take the limit as $C_M$ becomes much larger than $E$. In this idealized situation, the boundary layer mixes perfectly with the farm flow, and the performance is limited only by the dynamics at the interface between the boundary layer and the outer flow. The limiting value of $c_{fp}$ is then:
\begin{equation}
 \max_{c_{ft}', C_M} c_{fp} = \frac{8}{27} E.
\end{equation}

\paolo{If $E \simeq 0.16$, the maximum $c_{fp}$ is approximately $0.047$, which is an order of magnitude larger than the output of current wind farms, which is around $c_{fp} \simeq 0.005$} (as can be verified, for example, by inspecting values in Fig.~\ref{fig:stab}).

\begin{figure}[t]
\centering
	\includegraphics[width = 0.99\linewidth]{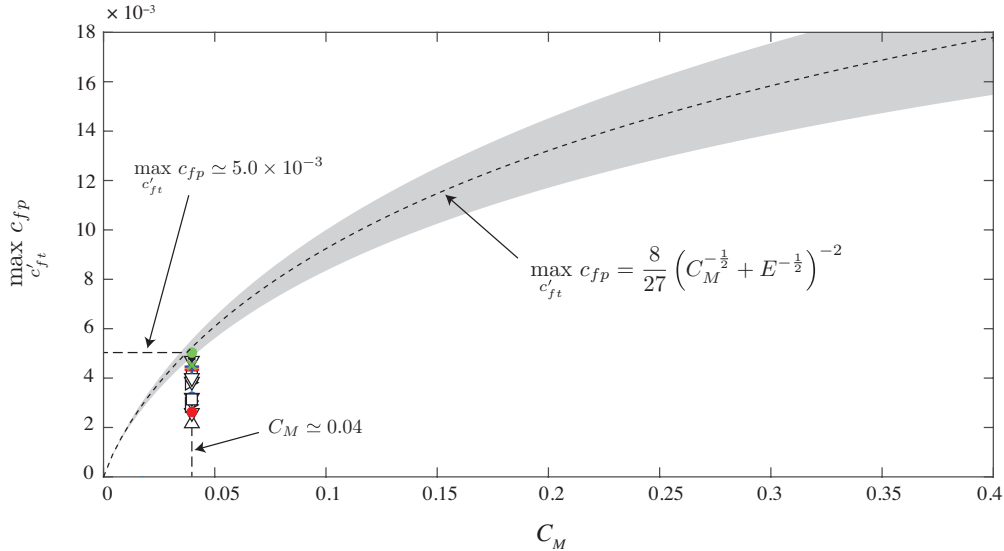}
	\caption{\paolo{Farm power coefficient $c_{fp}$ in a fully-developed wind farm, maximized over $c_{ft}'$, as a function of momentum transfer coefficient $C_M$. Symbols show available field data for existing wind farms (for which we assume $C_M = E/4 \simeq 0.04$). The dashed lines corresponds to our theory (with $E = 0.16$). The shaded region shows the effect of a $\pm 20\%$ change in $E$. Symbols as in Fig.~\ref{fig:Cfarm} and Table~\ref{table:data}; these symbols are displayed to show that LES, experiments and field measurements are consistent with the upper bound proposed here. 
		}
		} \label{fig:Cfarm_vs_CM}
\end{figure}

In practice, however, reaching this limit requires $C_M^{1/2} \gg  E^{1/2}$, implying $C_M$ would need to be of order ten, which is several hundred times current values (which we estimated around 0.04 in Sec.~\ref{sec:getCM}). Nevertheless, it is instructive to consider the change in $c_{fp}$ that would be accomplished by increasing $C_M$ through an order of magnitude, as plotted in Fig.~\ref{fig:Cfarm_vs_CM}. The regime accessed by current wind farms corresponds to $C_M \simeq 0.04$, $c_{fp} \simeq 0.005$. Our theory predicts, for example, that increasing $C_M$ to 0.4 would yield a $c_{fp}$ of approximately 0.018, corresponding to over three times its current value. This indicates that even moderate increases in the momentum exchange coefficient $C_M$ could lead to practically significant gains in turbine performance.

\section{Discussion}\label{sec:disc}
Our model emphasizes turbulent {and dispersive} transport at two planar interfaces as the mechanism enabling energy extraction by the array. This appears to be a reasonable assumption in the fully developed regime; however, it would not hold near the front of a turbine array, where the turbine wakes are separated by regions of relatively clean air. This issue would likely result in strong disagreements between predicted and measured power outputs (which should scale with the average of the cube of $u$, rather than the cube of the average, which we used in defining $c_{fp}$). 

{A point worth noting is that the wind farm boundary layer is relatively thin; for example, for the Lillgrund wind farm (which has the fastest-growing boundary layer, on account of its larger $c_{ft}'$), $dh_b/dx$ from our model is only \paolo{0.042}. If we consider flow along the longest direction of the Lillgrund farm, we can estimate, at the end of the farm, $h_b \sim h_f$, implying that the top of the boundary layer reaches a height of the order $\sim 2h_f$. Consistently with this estimate, a recent simulation of the complete Lillgrund wind farm  \cite{Nilsson_etal_WE_2015} has obtained excellent agreement with field measurements while using a computational domain having an overall vertical height of only $2.5D \simeq 2.1h_f$.}

\begin{figure}[t]
\centering
	\includegraphics[width = 0.99\textwidth]{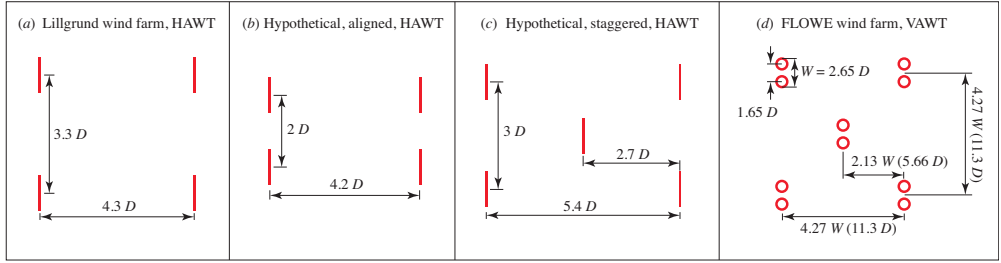}
    \vspace{10pt}
	\caption{{Illustrations of turbines in dense layouts. ($a$) Lillgrund wind farm (comprising horizontal-axis wind turbines) \cite{Dahlberg_Thor_Report_2009}. The next sketches show hypothetical ($b$) aligned and ($c$) staggered configurations that would provide an order of magnitude increase in power density, provided that a nearly complete wake recovery were to be achieved between turbine pairs. These are only slightly more closely spaced than existing large wind farms. ($d$) shows a sketch of the recently developed FLOWE facility (consisting of pairs of vertical-axis turbines) \cite{Kinzel_etal_JOT_2012}.}} \label{fig:layout}
\end{figure}

{According to Table~\ref{table:data}, removing wind shadowing effects between turbines (such that all turbines would output as much power as the first one) in a wind farm with dense spacing such as Lillgrund (sketched in Fig.~\ref{fig:layout}$a$) would yield a power output that is roughly 3.3 times that of existing arrays (in the fully developed regime).}
This is in contrast with the ten-fold improvement identified by the ideal limit in Sec.~\ref{sec:limit}. This ideal limit proposes that the array be redesigned to enhance the downward transport of momentum, relative to a traditional boundary layer (which constrains the operation of even a single turbine). 
As noted in Sec.~\ref{sec:limit}, approaching the ideal wind farm limit requires a $C_M$ of around 10, which may not be achievable; it would be interesting to investigate bounds for the value of $C_M$, as this would help establishing a sharper ideal wind farm limit. {Nevertheless, it is interesting to note that, if one could hypothetically achieve wake recovery over a length of approximately $4D$, the order-of-magnitude improvement in power density could be obtained by turbine layouts that are only slightly denser than for the Lillgrund array, as sketched in Fig.~\ref{fig:layout}($b,c$).}

We should briefly also comment here on recent works  \cite{Dabiri_JRSE_2011, Kinzel_etal_JOT_2012} involving bio-inspired layouts of counter-rotating, vertical-axis turbines {(sketched in Fig.~\ref{fig:layout}$c$)}, which have reported increases in power density of one order of magnitude, relative to the fully-developed regime of {Horns Rev} and other large wind farms. Our model suggests that such an order-of-magnitude increase in power density might be difficult to achieve in the fully developed regime. Concerning this issue, it would be interesting to examine in detail the contribution of developing-flow effects to the large power output described in Refs.~\cite{Dabiri_JRSE_2011,Kinzel_etal_JOT_2012}. 
It remains of great practical interest to develop simple and accurate theoretical models of developing flows in wind farms and other canopies. It would also be valuable to construct a general framework linking the kinematics of an array of obstacles (such as the  rotating-cylinder configurations recently examined in Refs.~\cite{Craig:2016dw,Craig:2016dn}) to a predicted value of the momentum exchange coefficient $C_M$, and therefore to achievable power output.

\section{Conclusions}\label{sec:conclusions}
In this paper, we introduce an entrainment-based model for the power output from the fully-developed region of large wind farms. The model comprises a fully-developed canopy whose upper edge is bounded by a self-similar boundary layer, and can represent arrays of turbines with arbitrary design and layout. We obtain a concise closed-form expression for the power density, which is in agreement with available field data, \paolo{experiments, and large-eddy-simulations}.
Our theory ultimately frames the performance of a large wind farm in terms of two parameters associated with turbulent and dispersive transport, namely $E$ (the entrainment coefficient at the edge of the boundary layer) and $C_M$ (the momentum exchange coefficient at the wind farm-boundary layer interface). A valuable feature of this approach is that it enables the use of existing parameterizations from the study of geophysical flows, and from canonical studies of turbulent flows. 

By introducing equations governing the potential temperature in the farm and in the boundary layer, we extend our theory to describe the effect of atmospheric stability on power output, finding agreement with the majority of available field measurements (with the exception of very stable conditions, where the theory underpredicts the power output). The theory's predictions for very stable conditions might be improved by new parameterizations that account for the effect of strong bottom roughness in stratified flows.
We build on our model and define an ideal performance limit, for a large wind farm, by considering a turbine array that has been optimized to maximise energy exchanges at the interface between the array and the boundary layer (such that $C_M$ is relatively large). In this idealised situation, the power output is limited by the rate at which energy is entrained at the upper edge of the boundary layer. The limiting value of power density (normalized by the power flux far above the wind farm) is $8E/27$, where $E$ is the entrainment coefficient (typically of order 0.1 in high-$Re$, unstratified flows). This simple result constitutes a reference value against which one can measure the performance of existing and proposed wind farm designs, and corresponds to an order of magnitude increase relative to current turbine arrays. 

In practice, reaching this ideal $8E/27$ limit requires an increase in $C_M$ by a factor of 400. Nevertheless, even moderate increases in $C_M$ could have a significant impact; for example, our theory predicts that increasing $C_M$ by 20\% (above its current value) would yield a corresponding 13\% increase in power output.
Our work points to a new {system-wide} approach for increasing wind farm performance, {focusing} on maximizing the momentum exchange coefficient $C_M$ {for the whole farm}, rather than on optimizing individual turbine performance. We suggest here that this may be achieved by redesigning wind farms to control the large-scale turbulent structures governing momentum fluxes. This is in direct contrast with current practice, which relies on optimizing the layout of turbines, each of which is designed in isolation. While it must be emphasized that the link between power density and electricity cost can be complex \cite{Meyers_Meneveau_WE_2012}, we are hopeful that this approach might enable marked decreases in the unit price of wind-generated electricity.

\section*{Acknowledgments}
P. L.-F. gratefully acknowledges support from a Junior Research Fellowship from Churchill College, Cambridge. {We are {also} grateful to Claudia Cenedese for sharing with us the data used to construct the entrainment parameterization, to Heidi Nepf for sharing data on canopy measurements, {and} to Antonio Segalini for helpful comments on a draft of this manuscript.}



\end{document}